\documentclass[aps,twocolumn,showpacs,preprintnumbers,amsmath,amssymb] 
{revtex4}

\newcommand{\beq}{\begin{equation}}
\newcommand{\eeq}{\end{equation}}

\usepackage{graphicx}
\usepackage{amsfonts}
\usepackage{epsfig}

\begin{document}


\title{Black hole thermodynamics from simulations of lattice Yang-Mills theory}

\author{Simon Catterall$^1$}
\author{Toby Wiseman$^2$}

\affiliation{$^1$Department of Physics, Syracuse University, Syracuse,
NY13244, USA}
\affiliation{$^2$Theoretical Physics, Blackett Laboratory, Imperial
College London, London, SW7 2AZ, UK}

\date{March 2008}

\begin{abstract}
We report on lattice simulations of 16 supercharge $SU(N)$ Yang-Mills quantum mechanics in the 't Hooft limit.
Maldacena duality conjectures that in this limit the theory is dual
to IIA string theory, and in particular that the behavior of the
thermal theory at low temperature is equivalent to that of certain
black holes in IIA supergravity. Our simulations
probe the low temperature regime for $N \le 5$ and the intermediate and high temperature regimes for
$N\le 12$. We observe 't Hooft scaling and at low
temperatures our results are consistent with the dual black hole prediction. 
The intermediate temperature range is dual
to the Horowitz-Polchinski correspondence region, and our results are consistent with smooth behavior there. We include the Pfaffian phase arising from the fermions in our calculations where appropriate.
\end{abstract}

%
%

\maketitle

\vspace{-0.5cm}
%
\section{Introduction}
%
\vspace{-0.25cm}

String theory has provided remarkable insight into the quantum
physics underlying black holes. Much recent progress stems
from conjectured dualities, which, in an appropriate limit, relate the
finite temperature low energy supergravity
limit of the string theory to strongly coupled thermal field theory. The  
entropy of
the black holes that arise
in these supergravity theories can then be computed in principle by  
counting microstates in their dual field theories. The pioneering 
calculations of
black hole entropy in \cite{StromVafa, Strom} are examples where
the dual field theory is a 2-d conformal field
theory which allows computation of the entropy  
despite the strong coupling.

For a large number $N$ of coincident D$p$-branes in the `decoupling' limit \cite
{Mald,Itzhaki}
the dual field theory is $(1+p)$-dimensional strongly coupled
maximally supersymmetric $SU(N)$ Yang-Mills theory, taken in the 't Hooft limit. The case of 
D3-branes yields the AdS--CFT correspondence.
Analytic calculation of the  
corresponding black hole entropy of these
theories has proven elusive despite interesting attempts
\cite{MaldMinw}.

Here we use lattice methods to study
the thermal gauge theory and hence test these conjectured
dualities. The simplest case for lattice work corresponds
to D0-branes \cite{four},
where the dual is thermal
16 supercharge Yang-Mills quantum mechanics (the `BFSS model' \cite{BFSS}).
This theory has recently been numerically studied using
a non-lattice formulation \cite{Nishimura4,Nishimura16}.
Earlier analytic approaches used a variational method \cite{Kabat1,  
Kabat2}. Related zero temperature numerical works are \cite
{Campostrini1, Hiller, Lunin}.

In this letter we simulate the super quantum mechanics in the t'Hooft
limit over a range of temperature and present preliminary results.
We obtain intermediate temperature
results for $N \le 12$ and low temperature results for $N \le 5$.
We pay particular attention to the
continuum limit and the behavior of the important Pfaffian phase
arising from the fermions. More details of the  method and results will be given in \cite{long}.

\vspace{-0.5cm}
%
\section{Duality and black holes}
%
\vspace{-0.25cm}

The type IIA string theory reduces to a supergravity theory for  
low energies
compared to the string scale $(\alpha^\prime)^{-1/2}$. In this limit
the thermal theory contains black holes with $N$ units of D0-charge. Their energy, $E$, is a function of their Hawking temperature, $T$.
Defining $\lambda=N g_s\alpha^{\prime -3/2}$ where $g_s$ is the
string coupling, we may write a dimensionless energy and temperature $ 
\epsilon = E \lambda^{-1/3}$ and $t = T \lambda^{-1/3}$. One finds  
provided we take $N$ large and $t \ll 1$ the black hole is weakly  
curved on string scales and the quantum string corrections are  
suppressed. The energy of this black hole can
be precisely computed by standard methods \cite{Itzhaki} giving,
\begin{eqnarray}
\epsilon = c\; N^2 t^{14/5} \qquad c = \left( \frac{2^{21} 3^{12} 5^
{2}}{7^{19}} \pi^{14} \right)^{1/5} \simeq 7.41 .
\end{eqnarray}
Duality posits that the thermodynamics of this black hole
should be reproduced by the dual Yang-Mills quantum mechanics
at the same temperature with $g_s\alpha^{\prime -3/2} = g_{YM}^2$
so $\lambda$ is identified with the 't Hooft coupling.

In the large $N$ limit, at high temperatures $t > 1$, the bound
state of D0-branes is of order the size of the string scale, and hence
all $\alpha^\prime$ corrections are important. One should best think of the
configuration dominating the partition function as a hot gas of D0-branes bound by strings. Horowitz and Polchinski have argued that
the low temperature black hole and
high temperature gas are the asymptotic descriptions and intermediate
temperatures smoothly interpolate between these
\cite{HorowitzP}.

\vspace{-0.5cm}
%
\section{Lattice implementation}
%
\vspace{-0.25cm}

The 16 supercharge $SU(N)$ Yang-Mills quantum
mechanics arises from dimensional
reduction of $\mathcal{N} = 1$ super Yang-Mills in 10-d. The
10-d gauge field reduces to the 1-d gauge field $A$ and 9 scalars,
$X^i$, $i = 1,\ldots,9$ and the 10-d Majorana-Weyl fermion to
16 single component fermions, ${\Psi}_{\alpha}$, $\alpha=1,\ldots,16
$. All fields transform in the adjoint of the gauge group.
In order to simulate the theory we must integrate
out the fermions giving rise to a Pfaffian.  The continuum
Euclidean path integral, $Z = \int dA dX \mathrm{Pf} \left( \mathcal{O} \right) e^{ - S_ 
{bos}}$, is then given by,
\begin{eqnarray}
S_{bos} & = & \frac{N}{\lambda} \mathrm{Tr} \oint^R d\tau
        \left\{ \frac{1}{2} (D_{\tau} X_i)^2 - \frac{1}{4} \left[ X_i,
X_j \right]^2 \right\}  \nonumber \\
\mathcal{O} & = & \gamma^\tau D_{\tau}  - \gamma^i \left[ X_i, \cdot
\right]
\end{eqnarray}
The $\gamma^\tau, \gamma^i$ are the Euclidean Majorana-Weyl
gamma matrices, and we choose a representation where,
$\gamma^{\tau} = \left( \begin{matrix} 0 & \mathrm{Id}_8 \\ 
\mathrm{Id}_8 & 0 \end{matrix} \right)$. We take Euclidean time to have period $R$. 

We have a choice of fermion boundary conditions.
Thermal boundary conditions correspond to taking the
fermions antiperiodic on the Euclidean time circle and correspond
to a temperature $t=\lambda^{-1/3}/R$. We will also
employ periodic fermions and the continuum partition
function is then an index, with $t$ the inverse volume.
The Pfaffian is in general complex \cite{Nicolai}. It is important
in principle to include the phase of the Pfaffian in the Monte-Carlo  
simulation, and we discuss this later.

We discretize this continuum model as,
\begin{eqnarray}\label{eq:actiondis}
       S_{bos} &=&  \frac{NL^3}{\lambda R^3}  \sum_{a=0}^{L-1} \mathrm
{Tr}\left[ \frac{1}{2}  \left( D_+ X_i \right)_a^2 -\frac{1}{4} [ X_{i,a}, X_{j,a} ]
^2  \right] \nonumber \\
\mathcal{O}_{ab} & = & \left( \begin{matrix} 0 & (D_+)_{ab} \\ (D_-)_
{ab} & 0 \end{matrix} \right) - \gamma^i \left[ X_{i,a}, \cdot
\right]  \delta_{ab}
\end{eqnarray}
where we have rescaled the fields $X_{i,a}$ and $\Psi_{i,
\alpha}$
by powers of the lattice spacing $a=R/L$ where $L$ is the number of
lattice points to render them dimensionless.
 We have introduced a Wilson gauge link field $U_a$, and taken covariant difference operators 
$(D_- W)_a = W_{a} - U_a^{\dagger} W_{a-1} U_a$, $(D_+ W)_a = U_a W_{a+1} U_a^{\dagger} - W_{a}$. Notice that  
the fermionic operator is
free of doublers and is manifestly antisymmetric.
This lattice action is finite
in lattice perturbation theory and hence will flow {\it without fine
tuning} to the correct supersymmetric continuum theory as the lattice  
spacing is reduced \cite{four, long}.

We use the RHMC algorithm \cite{rhmc} to sample configurations using
the absolute value of the Pfaffian. The phase may be
re-incorporated in the expectation value of an observable ${\cal A}$
by reweighting as $<\mathcal{A}> = \frac{\sum_{m} \left( \mathcal{A}  
e^{i \phi}
\right)}{\sum_{m} \left( e^{i \phi} \right)}$.
Here $e^{i \phi(\mathcal{O})}$ is the phase of the Pfaffian
and the sum runs over all members of our phase quenched ensemble.

We find in practice that the RHMC simulation of the thermal theory at
low temperature, $t \lesssim 1$, exhibits an instability corresponding
to the scalar fields moving out along the flat directions of
the classical potential. Hence the algorithm never thermalizes and cannot be used to approximate the path integral. This has been observed before \cite
{Nishimura16}.
We believe this divergence is a lattice artifact that is related to
the discretization of the fermion operator. In previous work \cite
{four} we have simulated the 4 supercharge quantum mechanics over a
range of $t$, using a Weyl representation for the fermions where one
obtains a real positive determinant. However, we have also tried  
using a Majorana
representation where one obtains a Pfaffian which we have discretized
in analogy with the 16 supercharge case discussed here.
While no divergence of the scalars was observed
in the Weyl simulations over a
large range of $t$ \cite{four}, the Majorana implementation has the same
instability we observe in the 16 supercharge case for $t \lesssim 1$.
Since both representations
are equivalent in the continuum limit this implies that the
instability is not a property of the continuum theory as is claimed
in \cite{Nishimura16} but merely an artefact of finite lattice spacing. More details will be given in \cite{long}.

We find no such problem simulating the periodic theory at small $t$.
At low temperature we expect the thermal and periodic theory to
be similar, and the configurations that dominate the path integral
will be similar. Hence in order to simulate the thermal theory at low
temperature, $t  \lesssim 1$, we have employed a reweighting of the
periodic theory. We can expect to get good results for the thermal
theory at low temperature by computing expectation values using the
periodic theory, and reweighting as,
\begin{equation}
< \mathcal{A} >_T = \frac{ \sum_{m}^{(P)} \left( \mathcal{A} \;  
\mathrm{Pf}(\mathcal{O}
_T) / | \mathrm{Pf}(\mathcal{O}_P) | \right) }{ \sum_{m}^{(P)} \left 
( \mathrm{Pf}(\mathcal{O}
_T) / | \mathrm{Pf}(\mathcal{O}_P) |  \right)}
\end{equation}
where $\sum_{m}^{(P)}$ is a sum over the phase quenched ensemble  
generated for the periodic theory, $\mathcal{O}_P$ and $\mathcal{O}_T$ are the periodic and thermal fermion operators respectively, and $< \ldots >_T$ is the  
expectation value for the thermal theory.

\vspace{-0.5cm}
%
\section{Results}
%
\vspace{-0.25cm}

\begin{figure}[]
\begin{center}
\includegraphics[height=1.7in,width=3.1in]{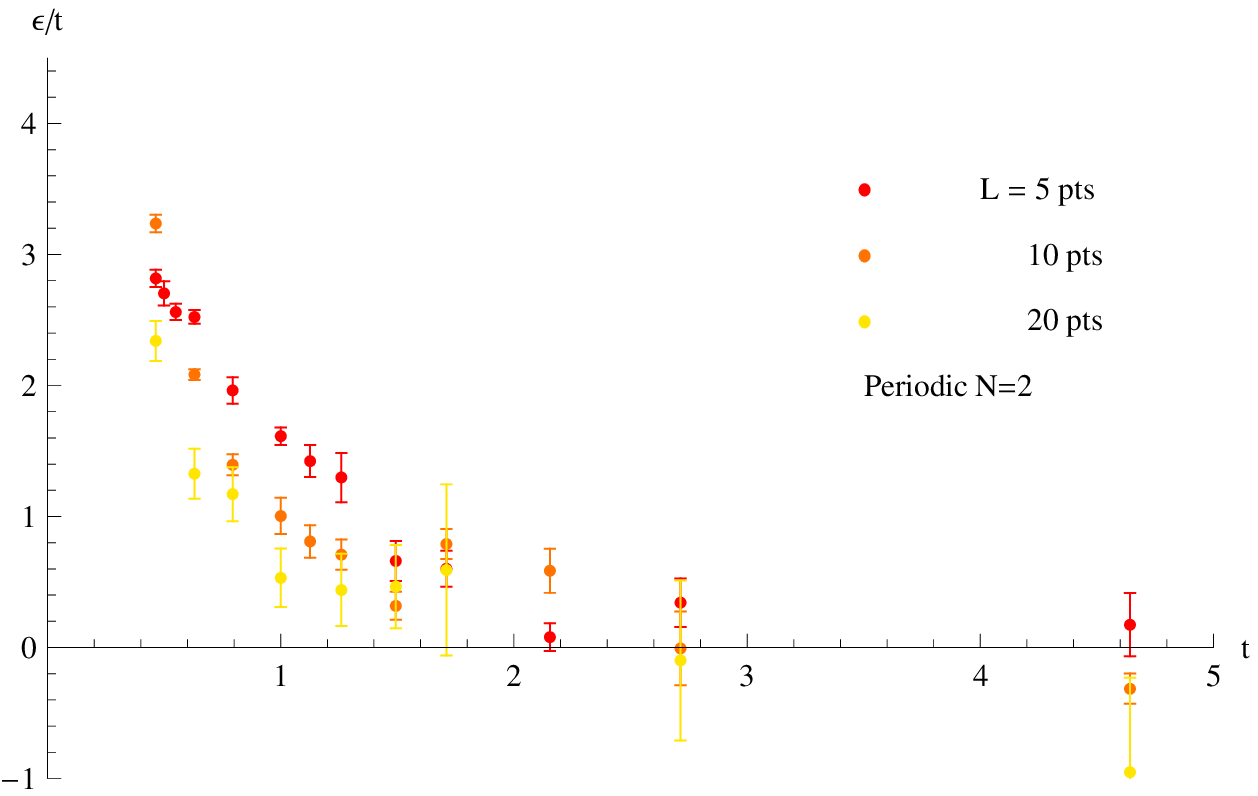}
\includegraphics[height=1.7in,width=3.1in]{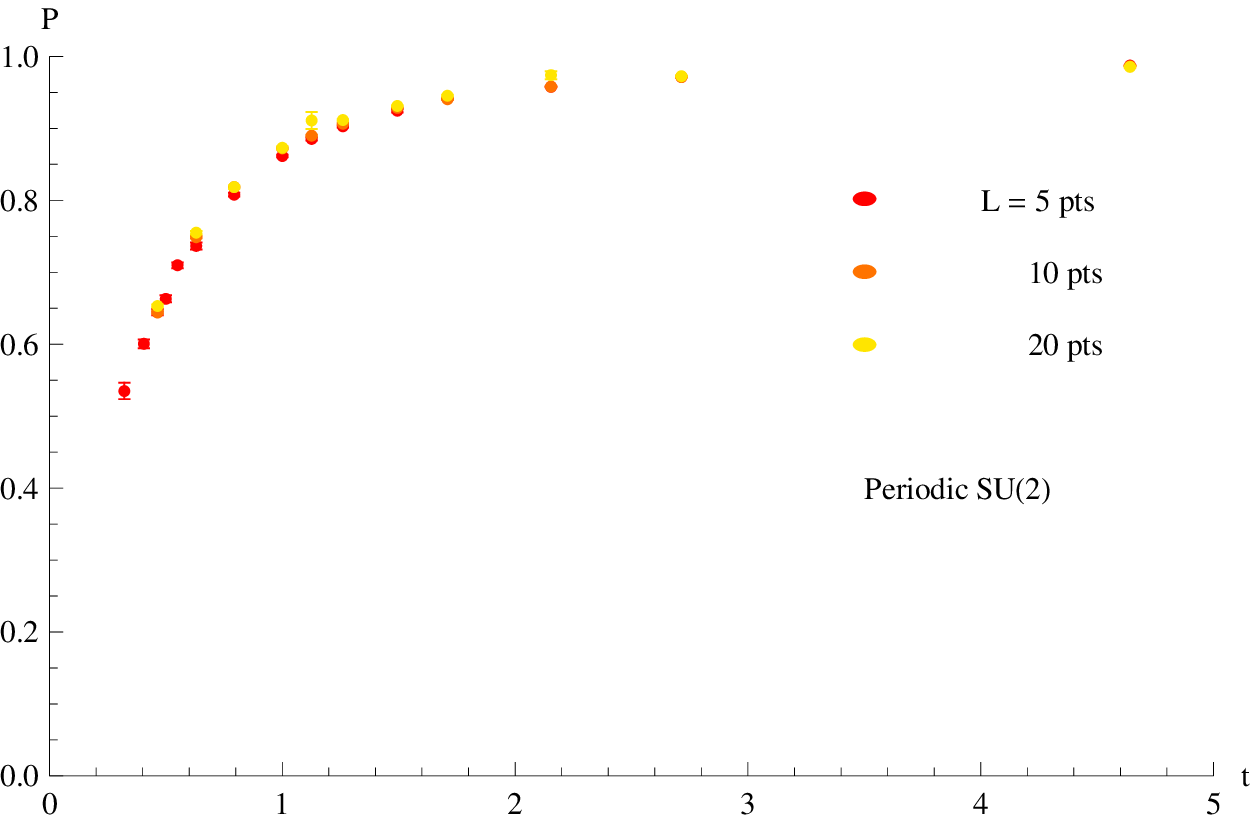}
\vspace{0.0in}
\caption{\emph{Top:} Plot showing $\epsilon/t$ verses dimensionless
temperature $t$ for the periodic $SU(2)$ theory for various numbers of
lattice points. \emph {Bottom:} Plot of the Polyakov loop against  
temperature for the same theory.} \vspace{-0.5cm}
\label{fig:susy}
\end{center}
\end{figure}

\begin{figure}[]
\begin{center}
\includegraphics[height=1.7in,width=3.1in]{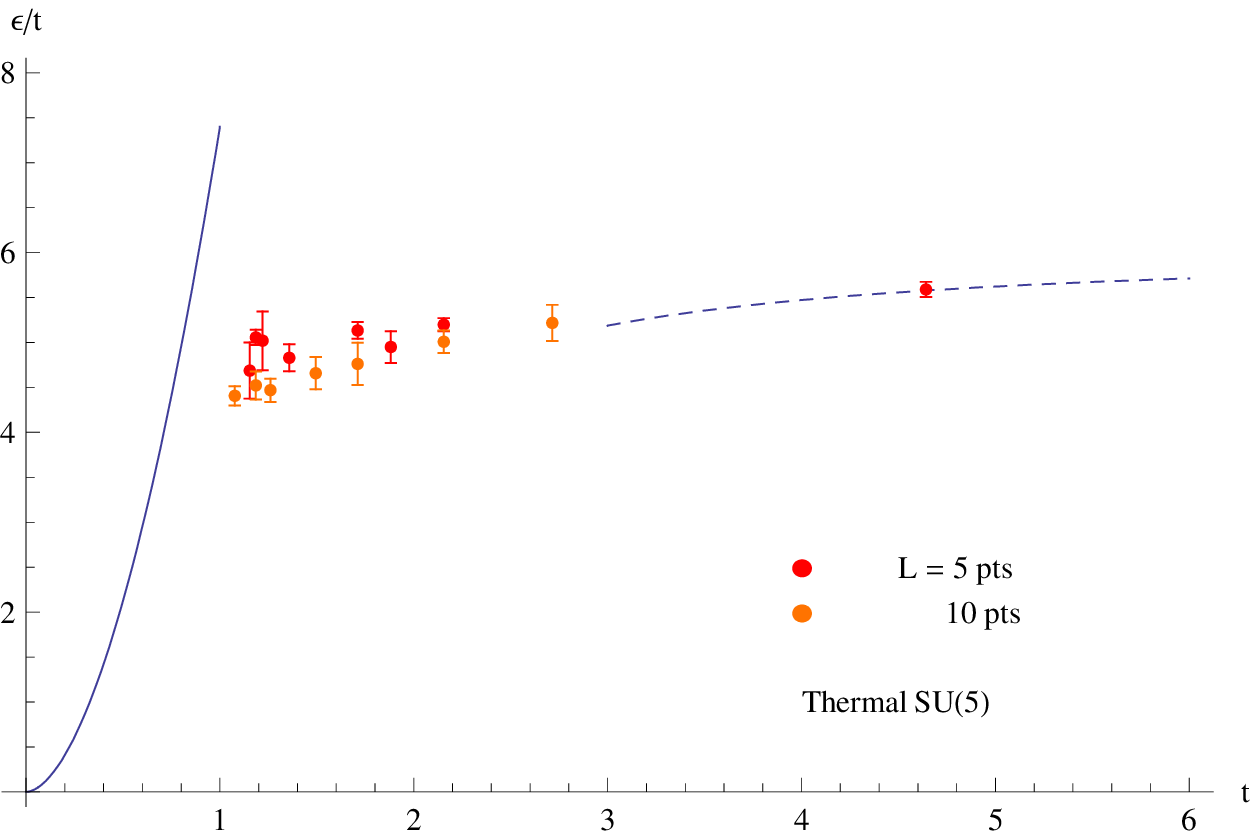}
\includegraphics[height=1.7in,width=3.1in]{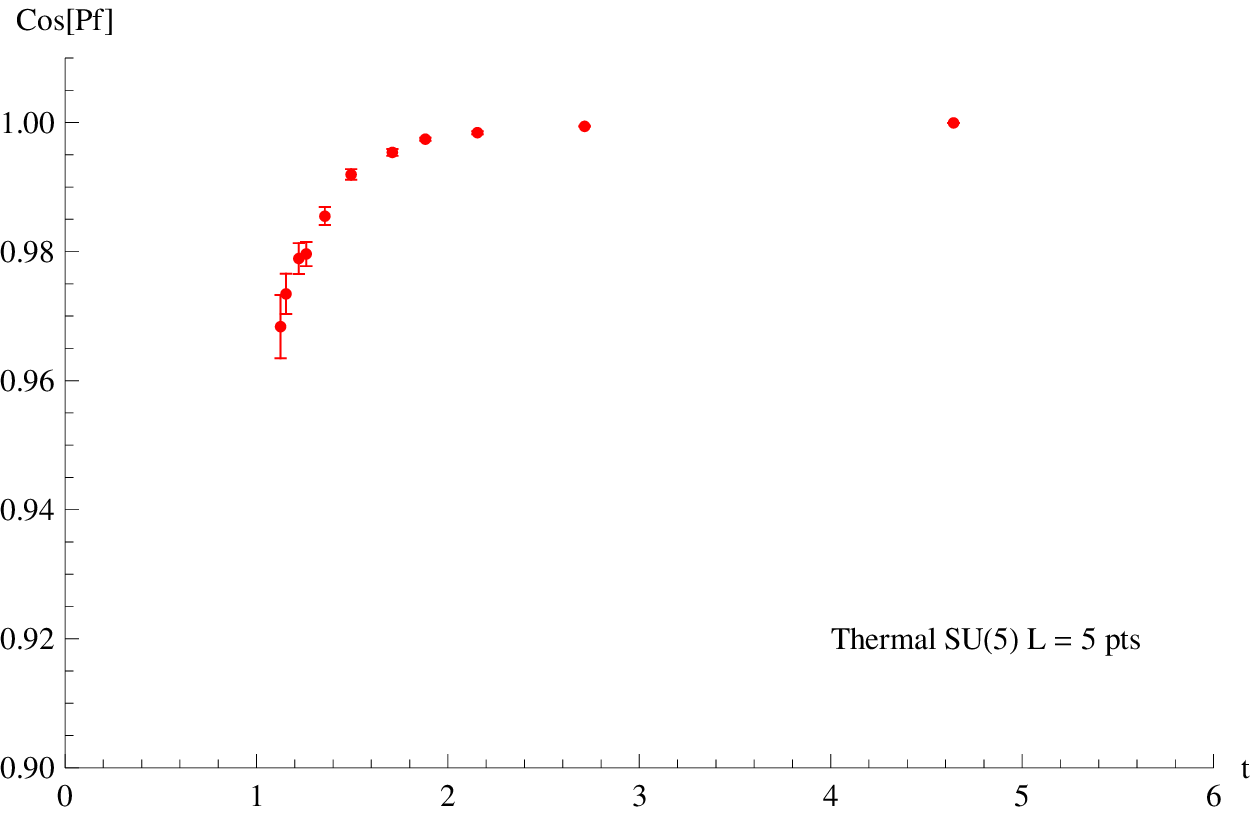}
\vspace{0.0in}
\caption{\emph{Top:} Plot of dimensionless energy $\epsilon/t$ verses
dimensionless temperature $t$ for the thermal $SU(5)$ theory with 5  
and 10 lattice points. \emph{Bottom:}
Plot of the cosine of the Pfaffian argument for the thermal
$SU(5)$ theory with 5 points. }  \vspace{-0.5cm}
\label{fig:contPfaff}
\end{center}
\end{figure}

\begin{figure}[]
\begin{center}
\includegraphics[height=2.in,width=3.1in]{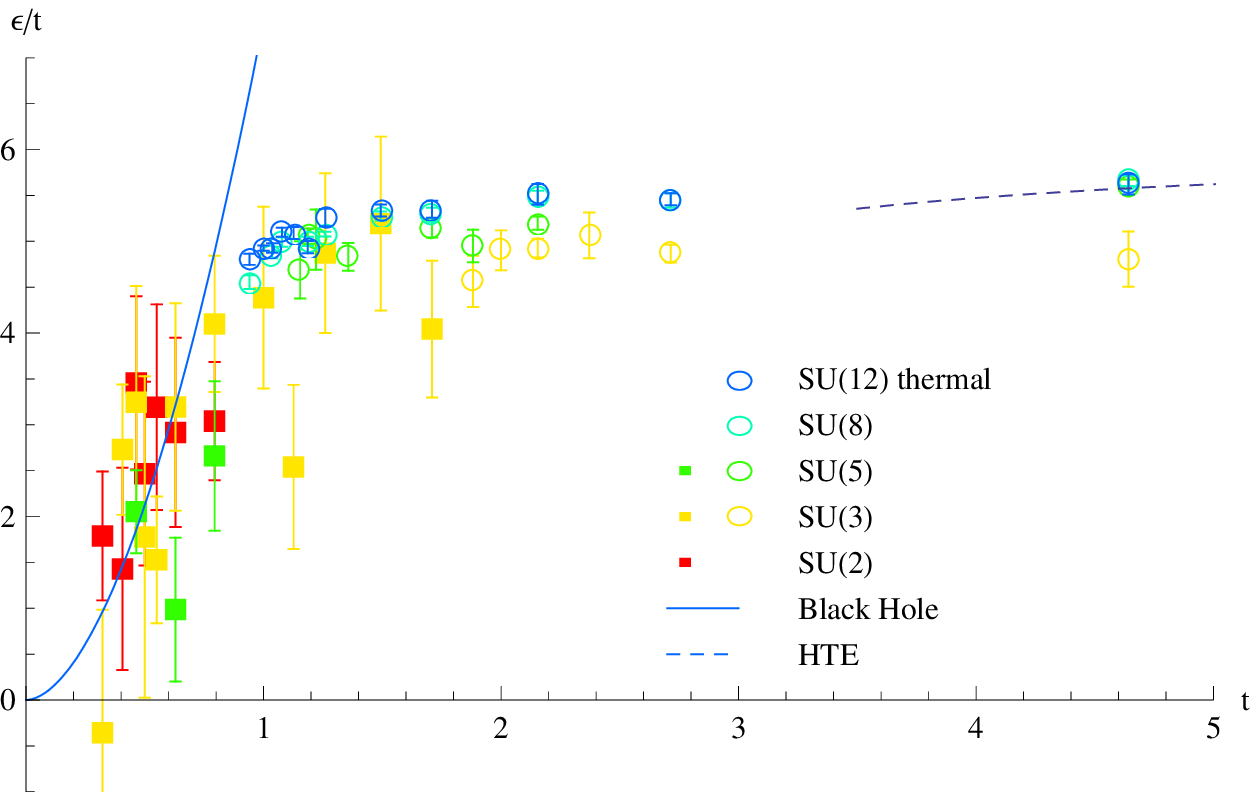}
\includegraphics[height=2.in,width=3.1in]{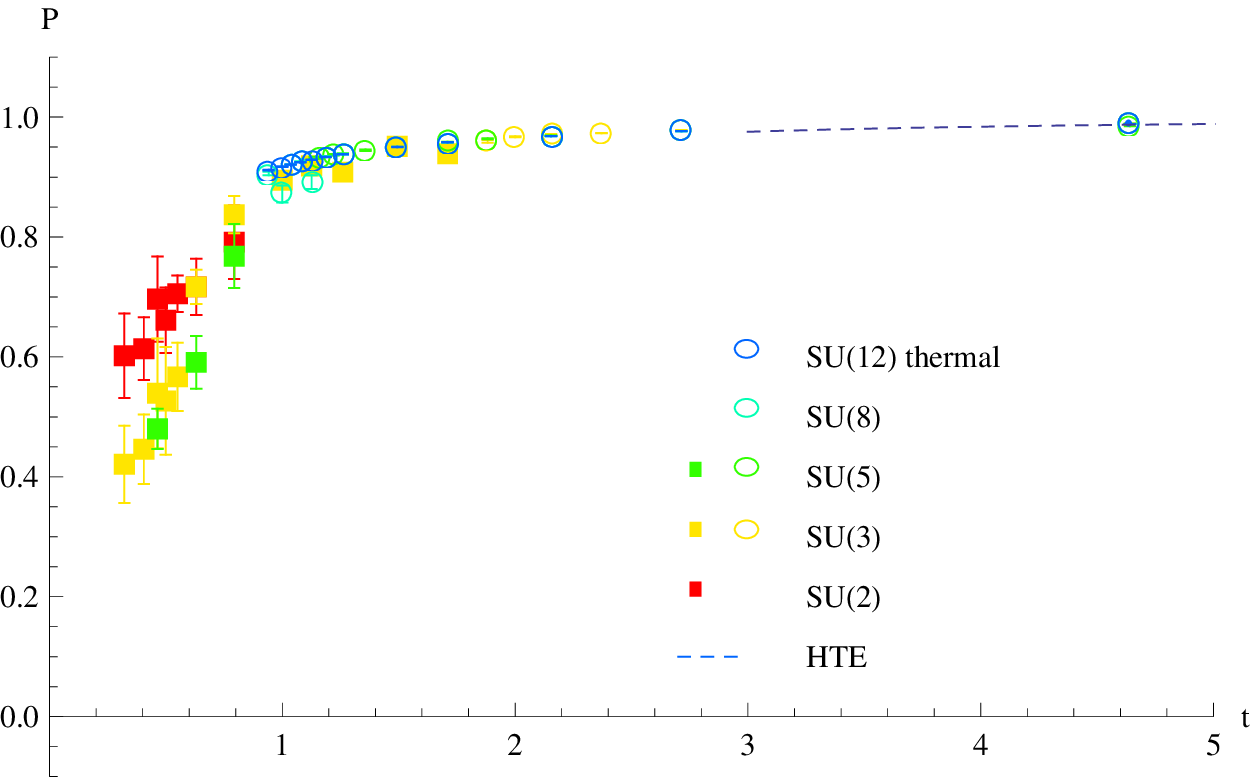}
\vspace{0.0in}
\caption{\emph{Top:} A plot of the dimensionless energy $\epsilon/t$
verses dimensionless temperature $t$. Data shown is generated
in two ways. For temperatures larger than $t \sim 1$ we simulate the
thermal theory for $N = 2,3,5,8,12$ with 5 points. The low
temperature results are computed for $N = 2, 3, 5$ for 5 points by
simulating the periodic theory, and reweighting with the appropriate
combination of the thermal and periodic Pfaffians, as described in
the main text. The low temperature black hole prediction is shown.
\emph{Bottom:} A plot of the Polyakov loop observable $P$ for the same  
cases.
}  \vspace{-0.5cm}
\label{fig:main}
\end{center}
\end{figure}

\begin{figure}[]
\begin{center}
\includegraphics[height=1.7in,width=3.1in]{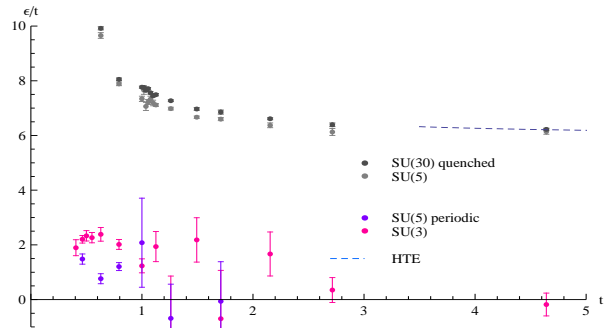}
\vspace{0.0in}
\caption{For comparison with figure \ref{fig:main}, $\epsilon/t$ verses $t$  
is shown for the quenched theory for $N = 5,12,30$ and periodic  
theory with Pfaffian reweighting for $N = 3,5$, using 5 point lattices.
}  \vspace{-0.5cm}
\label{fig:comparison}
\end{center}
\end{figure}

We have simulated the thermal and periodic theories concentrating on the
range $0.3 < t < 5$. We have focused on two observables - 
the mean energy, $\epsilon$, and absolute value of the trace of the 
Polyakov loop, $P$. In  
the Yang-Mills theory these are given by \cite{four,long}
\begin{eqnarray}<\epsilon /t> & = & \frac{3}{N^2}\left(\frac{9}{2} L (N^2-1)-
<S_{bos}>\right) \nonumber \\ 
P & = & \frac{1}{N}<|\mathrm{Tr}\prod_{a=0}^{L-1}U_a|> .
\end{eqnarray}
The inclusion of $1/N^2, 1/N$ in these definitions is
to ensure these quantities are finite in the t'Hooft limit for a deconfined phase. 
In the periodic case, since $Z$ is an index, it should not depend continuously on the inverse volume $t$, and hence in the continuum $\epsilon = 0$.

To check for a restoration of supersymmetry we have
computed $\epsilon/t$ in the
periodic theory for a variety of lattice sizes $L = 5, 10$ and $20$.
The upper
plot of figure \ref{fig:susy} shows $\epsilon/t$ for $SU(2)$.
For large $t$ the index $\epsilon$ is already consistent
with zero for $L=5$, while at small $t$ it appears to approach zero
as $L$ increases. Notice that while
this quantity is a sensitive test of the restoration of
supersymmetry in the continuum limit $L\to\infty$
other observables such as $P$ shown in the
lower plot
are relatively insensitive to the
number of lattice sites for $L \ge 5$.

We have also examined the continuum limit of the thermal theory. In
figure \ref{fig:contPfaff} we show $L = 5$ and $10$ data for the
thermal energy for $SU(5)$ (in the phase quenched
approximation -- which we discuss shortly). As noted above, we find a
lattice instability for the thermal theory with $t  \lesssim 1$ (with
some dependence on $N$ and $L$). However, for larger $t$ this does
not occur. As argued above we believe this is an artifact of our
lattice formulation and has nothing to do with continuum physics. The
points plotted in the figure are taken only from simulations where the
scalar distribution remained bounded for hundreds of physical RHMC
times (the observed instability sets in very quickly in RHMC time, so
the change in behavior is easy to identify). The plot shows  
that
these lattice spacing effects are small and hence for the remainder of  
our results
we show only data from $L=5$ point lattices.

In the lower plot of figure \ref{fig:contPfaff} we show
the mean cosine of the Pfaffian phase for the
thermal $SU(5)$ theory with $L=5$ lattice sites. As
expected this phase becomes more important at lower temperatures
but the actual value is close to
one over the range of temperatures where we can directly simulate the thermal
theory. Indeed the effects of reweighting are negligible in this temperature regime. Hence for the data we present later for direct simulation of the thermal theory we use the phase quenched approximation. Since the Pfaffian is very costly to compute this allows us to work at larger $N$.

We now turn to the main results of this letter.
In figure \ref{fig:main} we plot the energy
and Polyakov loop for various $N$ and $L=5$ point lattices.
For high temperature we have used direct simulation of the
thermal theory (phase quenched as discussed above),
and we are able to obtain results up
to $N = 12$.  At low temperatures we obtain results by a reweighting
of the periodic simulations as discussed above. The results
from both methods agree in the regime where they
overlap $t\sim 1$.

At very high temperatures the curves approach
a constant corresponding to the result from classical equipartition
assuming $N^2$ deconfined gluonic states (the fermions
are lifted out of the dynamics by their thermal mass in this
limit). In contrast for low temperatures the energy
approaches zero signaling the presence of a supersymmetric vacuum at
vanishing temperature.

As seen before \cite{four}, we see that t'Hooft scaling sets in for
small $N$, with $N=3$ already giving results close to an extrapolated
large $N$ result. The high temperature asymptotics computed in  \cite 
{NishimuraHTE} are also plotted
and agree with our data. Our results also
appear consistent with those found recently
using non-lattice methods \cite{Nishimura16}.

There are two important physical observations. Firstly
the curves appear to interpolate from high to low temperature
smoothly -- there is no obviously discontinuous behavior. This is to
be contrasted with the quenched version of this theory which
has a large $N$
confinement/deconfinement phase transition at 
$t \simeq 0.9$ \cite{Aharony1, Aharony2}. Since
the intermediate temperature range $t \sim 1$ is dual to the regime
where the thermal D0-branes have a radius comparable to the string
scale, we are probing the Horowitz-Polchinski
correspondence regime, and seeing apparently smooth behavior there.

Secondly, the low temperature behavior of the theory appears
consistent with the prediction from supergravity, also shown in the  
plots. This is to be contrasted with the quenched energy curve shown for  
comparison in figure \ref{fig:comparison} which departs strongly  
from the black hole prediction at low temperature. 
In this figure we also show 
the periodic theory which shows the degree of 
supersymmetry breaking for these lattices is small.

It would be very interesting
to extend these calculations to 2 and 3 dimensional Yang-Mills systems
which are thought to be dual to D1 and D2 brane systems using recent
lattice formulations retaining exact supersymmetry \cite{twisted}.

\vspace{-0.9cm}

%
\section*{Acknowledgements}
%

SC is supported in
part by DOE grant
DE-FG02-85ER40237. TW is supported by a PPARC advanced
fellowship and a Halliday award. Simulations were performed using
USQCD resources at Fermilab.

%
\bibliographystyle{apsrev}
\bibliography{ref}
%

\end{document}